\documentclass[preprint,prd,tightenlines,nofootinbib]{revtex4}

\pdfoutput=1
\usepackage{amsmath,bm}
\usepackage{graphicx}
\usepackage{hyperref} 

\usepackage{color}
\newcommand{\as}{\alpha_{\mathrm{s}}}
\newcommand{\LA}{\mathrm{A}}
\newcommand{\LB}{\mathrm{B}}
\newcommand{\LF}{\mathrm{F}}

\newcommand{\LS}{\mathrm{S}}

\newcommand{\LW}{\mathrm{W}}
\newcommand{\LZ}{\mathrm{Z}}

\newcommand{\Lb}{\mathrm{b}}

\newcommand{\Lg}{\mathrm{g}}

\newcommand{\Lt}{\mathrm{t}}

\newcommand{\tildenot}{{\raise.17ex\hbox{$\scriptstyle\mathtt{\sim}$}}} % "~"

\begin{document}

\title{Finding top quarks with shower deconstruction}

\author{Davison E. Soper}
\affiliation{
Institute of Theoretical Science\\
University of Oregon\\
Eugene, OR  97403-5203, USA\\
}

\author{Michael Spannowsky}
\affiliation{
Institute for Particle Physics Phenomenology\\
Department of Physics\\
Durham University\\
Durham CH1 3LE,
United Kingdom\\
}

\begin{abstract}
We develop a new method for tagging jets produced by hadronically decaying top quarks. The method is an application of shower deconstruction, a maximum information approach that was previously applied to identifying jets produced by Higgs bosons that decay to $\Lb\bar\Lb$. We tag an observed jet as a top jet based on a cut on a calculated variable $\chi$ that is an approximation to the ratio of the likelihood that a top jet would have the structure of the observed jet to the likelihood that a non-top QCD jet would have this structure. We find that the shower deconstruction based tagger can perform better in discriminating boosted top quark jets from QCD jets than other publicly available tagging algorithms.
\end{abstract}

\date{3 January 2013}

%\keywords{perturbative QCD, parton shower}

\preprint{IPPP/12/83}
\preprint{DCPT/12/166}

\maketitle
%-------------------------------------------------------------------
\section{Introduction}

A generic problem of some importance at hadron colliders like the Large Hadron Collider (LHC) is to find events generated by a signal process of interest among events generated by less interesting background processes. For this purpose, one often looks for events with one or more jets containing the decay products of a heavy particle that has been produced with a transverse momentum that is substantially larger than its mass, so that the sought decay products are part of a visible jet \cite{taggingef}. An important example is looking for jets that contain the decay products of a hadronically decaying top quark. One wants to distinguish top jets from the more numerous ordinary QCD jets that do not contain the decay products of a top quark. Experience shows that the analysis of jet substructure is useful for this purpose \cite{Chatrchyan:2012ku}.

Using jet substructure, one wants to be able to tag a jet with a label $t$ such that a jet with $t = {\tt top}$ is likely to be a top jet and a jet with $t = {\tt other}$ is not so likely to be a top jet. Several such top tagging algorithms are available \cite{brooijmans, Thaler:2008ju, JHtagger, Ellis:2009me, CMStagger, template, HEPtagger, Nsubjettinesstagger, notrees}. Somewhat more generally, one would like to be able to assign a real variable $\chi$ to a jet such that a large value of $\chi$ indicates a jet that is likely to be a top jet and a small value of $\chi$ indicates a jet that is unlikely to be a top jet. Then a ${\tt top}/{\tt other}$ tag can correspond to a cut on $\chi$, but the cut can be adjusted at will to increase or decrease the fraction of top jets that pass the cut while correspondingly decreasing or increasing the fraction of background jets that pass the cut.

In Ref.~\cite{SSI}, we described a method called shower deconstruction to distinguish signal jets from background jets. We applied the method to jets containing the decay products of a Higgs boson decay to ${\rm b} + \bar {\rm b}$. In this simple example, we found that shower deconstruction worked well enough to perform better than the Butterworth-Davison-Rubin-Salam (BDRS) method \cite{BDRS} in accomplishing the same end. In this paper, we extend the shower deconstruction method to finding top quark jets. This case includes richer physics: a) the top quark can decay but until it decays it can emit gluons and b) one of the daughter particles, the W boson, itself decays.

With this richer physics to work with, one might expect that shower deconstruction would do well compared to presently existing methods. To find out, we compute results for background fake rate versus signal acceptance obtained with shower deconstruction and compare to the results of existing top taggers.

Our plan is as follows. In Sec.~\ref{sec:deconstruction}, we very briefly describe the general ideas of shower deconstruction, referring to Ref.~\cite{SSI} for a fuller explanation of the method. In Sec.~\ref{sec:decays}, we describe in more detail the nature of a parton shower with decays and especially with decays of strongly interacting particles and with more than one level of decays. We concentrate on the physical principles and the main formulas and leave some details to an appendix \ref{sec:appendix}. Then in sections \ref{sec:results}, \ref{sec:resultslowPT} and \ref{sec:resultsdiffR}, we study the tagging performance of shower deconstruction, varying the boost of the possible top jet and the cone size used to define it. In Sec.~\ref{sec:Wmasshypothesis}, we explain how shower deconstruction could be used to measure a parameter of the signal theory, namely the W mass. Finally, in Sec.~\ref{sec:conclusion}, we offer some conclusions.

\section{Shower deconstruction}
\label{sec:deconstruction}

We seek to distinguish a jet that contains the decay products of a hadronically decaying top quark from a jet produced by ordinary QCD processes that do not involve a top quark. The jet to be examined is presumed to have a large transverse momentum, several hundred GeV. It is constructed with a standard jet algorithm, such as the Cambridge-Aachen algorithm \cite{CambridgeAachen}, using a large effective cone size so as to have a good chance of capturing the decay products of a top quark within the jet. This is the ``fat jet.''

We group the contents of the fat jet into narrow subjets, which we call microjets. In an experimental implementation, the microjets would be constructed directly from information on the energy deposits in the calorimeter and tracker, using as fine an angular resolution as is practical.  

The computational time needed to analyze an event increases quite quickly with the number of microjets. However, we find that the lowest transverse momentum microjets carry little useful information. Accordingly, we choose a number $N_{\rm max}$ with default value $N_{\rm max} = 9$ and discard the lowest transverse momentum microjets if there are more than $N_{\rm max}$ microjets, keeping the $N_{\rm max}$ microjets that have the highest transverse momenta. Additionally, we discard microjets with $p_{T}^{{\rm micro}} < p_{T,{\rm min}}^{\rm micro}$, with default value $p_{T,{\rm min}}^{\rm micro} = 5\ {\rm GeV}$.

This process gives the fine grained information with which we describe the fat jet in shower deconstruction: the four-momenta $\{p\}_N = \{p_1, p_2, \dots, p_N\}$ of the microjets. From these variables, we wish to construct a function $\chi(\{p\}_N)$ with the property that large $\chi$ corresponds to a high likelihood that the jet is a top jet. In fact, we define $\chi$ as the likelihood ratio
\begin{equation}
\label{eq:chidef}
\chi(\{p\}_N) = 
\frac{P(\{p\}_N|\LS)}{P(\{p\}_N|\LB)}
\;,
\end{equation}
where $P(\{p\}_N|\LS)$ is the probability density that a jet in a sample of top jets (``signal jets'') would have the configuration $\{p\}_N$ and $P(\{p\}_N|\LB)$ is the probability density that a jet in a sample of background jets  would have the configuration $\{p\}_N$. One might imagine constructing $P(\{p\}_N|\LS)$ and $P(\{p\}_N|\LB)$ by generating events with a trusted parton shower Monte Carlo event generator. However, that method is not practical. Instead, we calculate $P(\{p\}_N|\LS)$ and $P(\{p\}_N|\LB)$ by calculating the probabilities that a simplified approximation to a shower Monte Carlo event generator would generate $\{p\}_N$ according to the signal hypothesis and the background hypothesis, respectively. Our simplified approximation to a shower Monte Carlo event generator is based on the shower algorithms described in Refs.~\cite{NSI, NSII, NSIII, NScolor} and in unpublished work in this ongoing series of papers \cite{NSunpublished}. For a brief review of the structure of parton shower event generators, see Ref.~\cite{Gieseke:1900zz}.

\begin{figure}
\centerline{\includegraphics[width=12.0cm]{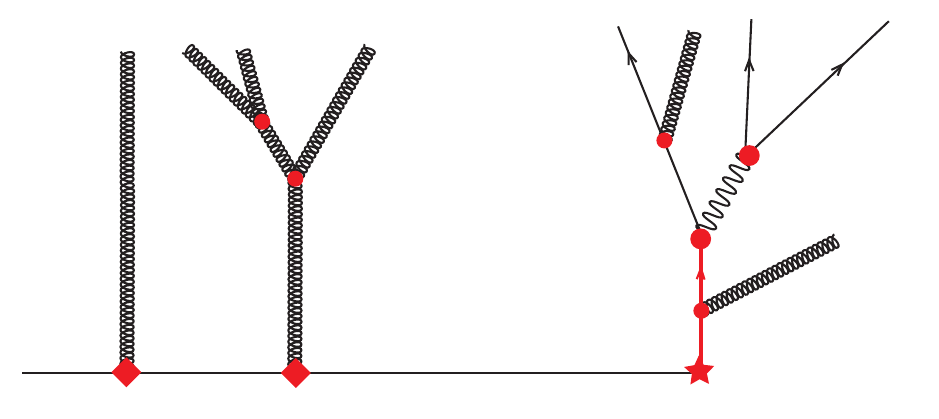}}
\caption{
Shower history for a top quark jet. The hard interaction is indicated by a star. Initial state emissions are indicated by diamonds. Parton decays are indicated by large filled circles and QCD splittings are indicated by small filled circles.
}
\label{fig:Shistory}
\end{figure}

How can one calculate these probabilities? Consider, for example, $P(\{p\}_N|\LS)$. We take $\{p\}_N$ to be the momenta carried by partons at a fairly late stage of a parton shower. In Fig.~\ref{fig:Shistory}, we show a possible shower history by which an event generator might generate a particular $\{p\}_N$. A top quark is created in a hard interaction, indicated by the star in the figure. In this shower history, the top quark emits a gluon. Then it decays into a W and a b quark. The b quark emits a gluon. The W decays to two light quarks. Meanwhile, initial state splittings, depicted by diamond vertices, create two gluons. After two QCD final state splittings, the two gluons have become four. Our shower model is simplified. Really, there are two incoming partons, ``a'' and ``b,'' that initiate the hard interaction. However, we do not distinguish which incoming partons split to create new partons. Also, we take the partons created by initial state splittings to be gluons.

%---------------FIGURE------------------
%-------------END FIGURE----------------

We should emphasize that not all partons in the event are represented in the shower history for the fat jet. One could depict a shower history for a whole event, but any parton in the complete shower history that does not have at least one descendant in the fat jet is left out of the shower history for the jet.

Now, given the shower history depicted in Fig.~\ref{fig:Shistory}, we assign a splitting probability or a decay probability to each vertex. The splitting probabilities are approximately the splitting probabilities that are used in parton shower event generators. They take into account information on color flow in the event history. The decay probabilities are approximately the decay probabilities that would be used in an event generator. Each propagator in the shower history corresponds to a Sudakov factor that gives, approximately, the probability not to have had a splitting between one vertex and the next or between the last vertex and the end of the shower. Thus, for a given shower history corresponding to the signal hypothesis, we calculate a probability density that that shower history would have produced the observed state $\{p\}_N$.

There are many shower histories that could lead to a given $\{p\}_N$. We sum the corresponding probabilities over all possible shower histories to calculate $P(\{p\}_N|\LS)$.

\begin{figure}
\centerline{\includegraphics[width=12.0cm]{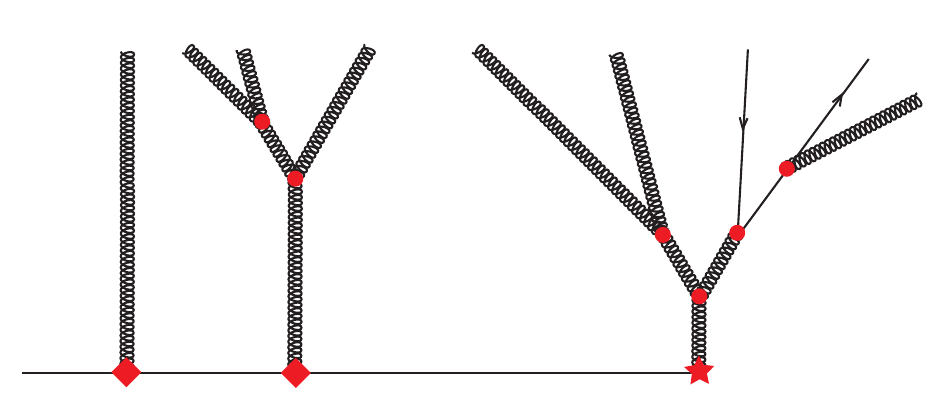}}
\caption{
Shower history for a QCD jet.
}
\label{fig:Bhistory}
\end{figure}

For the background hypothesis, we have different sorts of shower histories. One is shown in Fig.~\ref{fig:Bhistory}. Again, we calculate the approximate probability density that the shower history would have produced the observed state $\{p\}_N$. Then we sum the corresponding probabilities over all possible shower histories to calculate $P(\{p\}_N|\LB)$.

%---------------FIGURE------------------
%-------------END FIGURE----------------

Of course, this brief description leaves out a lot of details. Most of them are presented in Ref.~\cite{SSI}. Because they are of some importance to the structure of the model, we reiterate in Sec.~\ref{sec:kinematics} some specifics of the kinematics and the choice of shower time. Then, in Sec.~\ref{sec:decays}, we address some issues that arise with particles that decay, particularly with particles that carry color and decay.

\subsection{Kinematics and choice of shower time}
\label{sec:kinematics}

Each parton in a shower history carries a label. We denote the momentum of parton $i$ by $p_i$. The absolute value of its transverse momentum is $k_i$; its rapidity is $y_i$; its azimuthal angle around the beam axis is $\phi_i$; and its virtuality is $\mu_i^2 = p_i^2 - m_i^2$.

In this study, we take gluons, light quarks, and b quarks to have mass zero. This is not right for b quarks, but it should be a reasonable approximation as long as the b quark has a transverse momentum $k_i$ with $k_i \gg m_\Lb$. In the signal process, we also have top quarks and a W boson. These have a mass $m_\Lt$ and $m_\LW$ respectively.

For shower deconstruction, the momentum $p_J$ of a mother parton is related to the momenta $p_A$ and $p_B$ of the daughter partons by $p_J = p_A + p_B$. This is different from what an ordinary parton shower event generator does. In an ordinary parton shower event generator, $p_A$ is the sum of the momenta of its daughter particles, but modified to put it on shell, so $p_A^2 = 0$ for a massless parton. This modification is an approximation, imposed because the generator does not ``know'' what $p_A^2$ is at the time that parton A is generated in $J \to A + B$. Thus the best that the generator can do is to put parton A on shell. Then when parton A splits the event generator ``finds out'' what $p_A^2$ should be and takes the needed extra momentum from somewhere else in the event. For shower deconstruction, however, we do not need to make this approximation and, in fact, the relation $p_J = p_A + p_B$ is quite convenient. 

In each splitting function, there is a factor $1/\mu_J^2$ where $\mu_J^2$ is the jet virtuality, defined by
\begin{equation}
\label{eq:virtualitydef}
\mu_J^2 = (p_A + p_B)^2 - m_J^2
\;.
\end{equation}
Here $m_J$ is the top quark mass in the case that $J$ represents a top quark and otherwise $m_J = 0$. In calculating $(p_A + p_B)^2$, we do not approximate $p_A$ and $p_B$ as being on shell.

Parton splittings in the shower are ordered from hard to soft. Consider the splitting of parton $J$ with momentum $p_J$ and absolute value of transverse momentum $k_J$. A convenient way to do this is to assign to each splitting a shower time $t$,
\begin{equation}
\label{eq:showertimedef}
e^{-t} = \frac{\mu_J^2}{|Q_0| k_J}
\;.
\end{equation}
We divide the virtuality $\mu_J^2$ by the transverse momentum $k_J$ of the mother parton and, in order to make $\exp(-t)$ dimensionless, by a reference scale $|Q_0|$ on the order of the momentum transfer in the hard scattering that initiates the fat jet. The shower splittings are ordered in order of increasing $t$. This is the choice of ordering given in Eq.~(49) of Ref.~\cite{SSI}. It has the property of ordering splittings from hard to soft: $t \to \infty$ for any splitting that becomes infinitely collinear or infinitely soft.\footnote{There are a number of choices of ordering parameter that have this property. Our particular choice follows that made in from Ref.~\cite{NSunpublished}.}

In the case of a parton decay $J \to A + B$ rather than a splitting, we assign to the decay the shower time
\begin{equation}
\label{eq:showertimedecaydef}
e^{-t} = \frac{|p_J^2 - m_J^2|}{|Q_0| k_J}
\;.
\end{equation}
The only difference here is that $p_J^2$ can be less than $m_J^2$, so we use the absolute value of $p_J^2 - m_J^2$.

\section{Decaying particles}
\label{sec:decays}

The shower histories for the signal process considered have three stages, the first arising from the creation of the top quark, the others arising from the decays of the top quark and then of the W boson created in the top quark decay. The description of these stages is implicit in parton showers generally. Specifically, we follow unpublished work \cite{NSunpublished} in the series \cite{NSI, NSII, NSIII, NScolor}. 

In the first stage, a top quark is created in a hard process. We look for a high transverse momentum jet that contains the top quark. If we are to have a top quark jet, the top quark transverse momentum $k_t$ must be larger than the top mass $m_\Lt$. We may place cuts that require $k_t \gg m_\Lt$. Thus in a parton shower picture the top quark has a potentially large virtuality to start with and can radiate gluons. This gives what we can call shower I: the top quark can radiate one or more gluons and create a full parton shower as the gluons split. In this first shower, radiation from the initial state partons can also occur and create partons with angles that place them as part of the fat jet.

For this first shower, we need splitting functions for quarks and gluons other than the top quark, possibly with the top quark serving as color connected partner. Thus we allow a large mass for the color connected partner. We also need a splitting function for the top quark, this time with a massless color connected partner.

The first shower is suspended when the top quark decays. This happens at a varying shower time corresponding roughly to $|\mu^2_t| = |p_t^2 - m_\Lt^2| \sim m_\Lt\Gamma_\Lt$. With our definition of shower time, the first shower is suspended at shower time
\begin{equation}
\label{eq:endofShowerI}
|Q_0| e^{-t_1} = \frac{|p_t^2 - m_\Lt^2|}{k_t} \sim \frac{m_\Lt\Gamma_\Lt}{k_t}
\;.
\end{equation}

Now a second shower, shower II, is created by the decay $\Lt \to \Lb + \LW$. The $\Lb$ quark can emit a gluon, initiating a shower. There is a minimum value for the starting shower time, $t_0^{\rm II}$, for shower II. This is determined by the maximum of
\begin{equation}
\label{eq:startofshowerII}
|Q_0| e^{-t_0^{\rm II}} = \frac{\mu_J^2}{k_J}
\;.
\end{equation}
Here $J$ is the bottom quark just after the decay and $\mu_J^2$ is the virtuality in the splitting of the bottom quark. Let us look at this using $+,-,\perp$ momentum components\footnote{We use $v^\pm = (v^0 \pm v^3)/\sqrt 2$.} in a frame in which the top quark before the decay has large $+$ momentum, much larger than the top mass, and zero transverse momentum. In this frame, the top quark momentum is approximately
\begin{equation}
p_t = \left(\sqrt 2 k_t, \frac{m_\Lt^2}{2\sqrt 2 k_t}, \bm 0 \right)
\;.
\end{equation}
The momentum of the bottom quark is
\begin{equation}
p_J = \left(\sqrt 2 z k_t, \frac{\mu_J^2 + \bm \kappa^2}{2\sqrt 2 z k_t}, \bm \kappa \right)
\;.
\end{equation}
Here $k_J = z k_t$. The momentum of the W boson is
\begin{equation}
p_\LW = \left(\sqrt 2 (1-z) k_t, \frac{m_\LW^2 + \bm \kappa^2}{2\sqrt 2 (1-z) k_t}, -\bm \kappa \right)
\;.
\end{equation}
Momentum conservation for the $-$ component of momentum gives
\begin{equation}
\frac{m_\Lt^2}{2\sqrt 2 k_t}
= \frac{\mu_J^2 + \bm \kappa^2}{2\sqrt 2 z k_t}
+ \frac{m_\LW^2 + \bm \kappa^2}{2\sqrt 2 (1-z) k_t}
\;.
\end{equation}
Thus
\begin{equation}
\frac{\mu_J^2}{k_J}
=
\frac{m_\Lt^2}{k_t}
- \frac{m_\LW^2}{(1-z) k_t}
- \frac{\bm \kappa^2}{z(1-z) k_t}
\;.
\end{equation}
This is maximized for $\bm \kappa = 0$ and then for $z = 0$. This gives
\begin{equation}
\frac{\mu_J^2}{k_J}
<
\frac{m_\Lt^2 - m_\LW^2}{k_t}
\;.
\end{equation}
Thus shower II starts at the starting splitting scale 
\begin{equation}
|Q_0| e^{-t_0^{\rm II}} = \frac{m_\Lt^2 - m_\LW^2}{k_t}
\;.
\end{equation}
Of course, this calculation has assumed that the top quark and the W boson are on shell. This is not exactly true in a real shower event, but it should be an adequate approximation.

In shower II, the bottom quark and its descendants can emit gluons, which can either be collinear to the mother parton or soft. One can also have initial state radiation of gluons: the top quark is the initial state parton whose decay starts shower II and it can radiate gluons just before the decay. Now, in shower II, the virtuality of a splitting is never large compared to $m_\Lt^2$. For that reason, there is never an approximate collinear singularity for gluon emission collinear with the top quark. However, there is a singularity corresponding to soft gluon emission. Recall that one can think of soft gluons as being emitted from color dipoles. Thus, in shower II, a soft gluon can be emitted from a dipole consisting of the top quark just before the decay and the bottom quark or one if its daughter partons. Normally, one would partition the splitting function for gluon emission from such a dipole into two terms, as we do for other dipoles. One term would correspond to gluon emission from the top quark and the other would correspond to gluon emission from the bottom quark or its daughter parton. However, it will be simpler for us not partition emissions from this dipole. We simply treat the gluon emissions kinematically as coming from the bottom quark or its descendants, with a splitting function that accounts for graphs in which the gluon is soft and connects with the top quark in the eikonal approximation.

Shower II is suspended at a splitting time corresponding to the W boson decay. This happens roughly when $|p_\LW^2 - m_\LW^2| \sim m_\LW\Gamma_\LW$. With our definition of shower time, the second shower is suspended at shower time around
\begin{equation}
\label{eq:endofShowerII}
|Q_0| e^{-t_2} \sim \frac{m_\LW\Gamma_\LW}{k_\LW}
\;.
\end{equation}

Now a third shower is created by the decay $\text W \to q + \bar q$. Either of the new quarks can emit a gluon, initiating a shower. Shower III starts at the starting splitting scale
\begin{equation}
|Q_0| e^{-t_0^{\rm III}} =
\left[\frac{\mu^2_J}{k_J}\right]_{\rm max}
= \frac{m_\LW^2}{k_\LW}
\;.
\end{equation}
(The derivation of this follows the derivation above for the start of shower II.)

What happens to the ``suspended'' showers? Let us suppose that $t_1 > t_2$. Then the second shower is suspended before it reaches shower time $t_1$. Now we start the third shower. When the third shower reaches shower time $t_2$, the partons in the third shower are splitting on a slow enough time scale that their splittings can interfere with splittings from the second shower. Thus we continue both of these showers together. We now have the possibility that partons in shower II can have partons in shower III as color connected partners and vice versa. However, this doesn't happen because the W boson carries no color, so that the partons in shower III are in any case color connected only to each other. Now the combined showers II and III continue until they reach shower time $t_1$. Then the partons in the combined showers II and III are splitting on a slow enough time scale that their splittings can interfere with splittings from shower I. Thus we continue all three showers together. We now have the possibility that any parton can be color connected to any other parton. The complete shower evolves until the end of showering. The method for restarting suspended showers is analogous if $t_2 > t_1$. 

We see that a parton shower that properly accounts for interference effects within the leading color approximation will reset color connected partners when one of the subshowers reaches the shower time at which a parent shower was suspended. This procedure will affect wide angle splittings at rather small virtualities. We expect that the effect of reseting color connections will not be numerically very significant. Thus in shower deconstruction in this paper, we omit the step of reseting color connections in this way.

If the original top quark has high enough transverse momentum, more top quarks can be created within shower I: the top quark can emit a gluon and the gluon can split into a $\Lt$-$\bar{\Lt}$ pair. This can happen more than once. Each $\Lt$ or $\bar{\Lt}$ thus created evolves until it is nearly on shell. Then each decays to $\Lb + \LW^+$ or $\bar{\Lb} + \LW^-$, creating a new independent bottom quark sub-shower. Each W also decays, creating a new independent subshower if it decays hadronically. At later stages, all of the subshowers rejoin. In the situation that we consider in this paper, the top quark transverse momentum is not large enough compared to $2 m_\Lt$ for these effects to be important, so we simply ignore the possibility of multiple $\Lt$-$\bar{\Lt}$ creation.

\section{Results for moderately boosted top jet}
\label{sec:results}

\subsection{Generating events}
\label{sec:EventSelection}

In order to test how well shower deconstruction works for finding top quark jets, we
generate signal $\text t \bar{\text t}$ and background dijet samples using standard QCD processes in \textsc{Pythia 8} \cite{Pythia} and \textsc{Herwig++} \cite{Herwig}. We remove the invisible particles from the fully hadronized final state and use the remaining particles with $|y|<5.0$ as input for the Cambridge-Aachen jet-finding algorithm \cite{CambridgeAachen} as implemented in \textsc{Fastjet} \cite{Cacciari:2005hq} with $R=1.0$. To accept an event we require at least two jets  with $p_{T,j}>500$ GeV each. We then analyzed the two jets with the highest $p_{T,j}$.

We analyze each fat jet using shower deconstruction. Additionally, we independently tag each of the jets as a top quark jet or not using four different taggers: the Johns Hopkins tagger \cite{JHtagger}, the CMS tagger \cite{CMStagger}, the HEPTopTagger \cite{HEPtagger}, the NSubjettiness tagger \cite{Nsubjettinesstagger}. These taggers take as input the individual hadrons that make up the fat jet. Shower deconstruction aims to take the finite resolution of the detector into account by operating on small reconstructed jets instead of hadrons. We call the small jets microjets. To construct the microjets, we use the $k_T$-algorithm \cite{kTjets} with $R=0.2$.

\subsection{Parameters for top taggers}
\label{sec:taggerparameters}

For shower deconstruction, we remove microjets from the analysis unless $p_{T}^{\rm{micro}}>5.0$ GeV. If more than nine microjets are left, we remove those with the lowest $P_T$ values until nine microjets remain.

Each of the top taggers other than shower deconstruction constructs a top mass and a W mass for each jet that meets the structural criteria of the tagger. Each tagger then requires that these reconstructed masses be in specified windows. We specify that a top is correctly reconstructed in a window\footnote{The measurement of the resonance's mass is subject to experimental limitations. We choose the mass windows large enough to reflect these limitations \cite{JHtagger,CMStagger, HEPtagger}.} of $172.3 \pm 25.0$ GeV and a W in a window of $80.4 \pm 10.0$ GeV, except for the NSubjettiness tagger where $160.0 \leq m_\Lt \leq 240$ GeV and $\tau_3/\tau_2 < 0.6$ as recommended in \cite{Nsubjettinesstagger}. 

Top tagging based on shower deconstruction uses the full decay matrix element including a Breit-Wigner factor to assign a weight for a given microjet configuration. Thus, the total widths of the top quark and the W boson are input parameters. However, because the physical widths are very small, we assume that the invariant mass of a set of microjets cannot be resolved at the level of the physical widths. To take these experimental uncertainties into account, we use values for the top width and the W width equal to half of the corresponding total mass window, {\it i.e.} $\Gamma_\Lt \to 25\ \text{GeV}$ and $\Gamma_\LW \to 10\ \text{GeV}$. We have checked that the performance of shower deconstruction is not highly sensitive to this choice.

Other parameters for shower deconstruction are as in Ref.~\cite{SSI}.

%---------------FIGURE------------------
\begin{figure}
\centerline{\includegraphics[width=8.0cm]{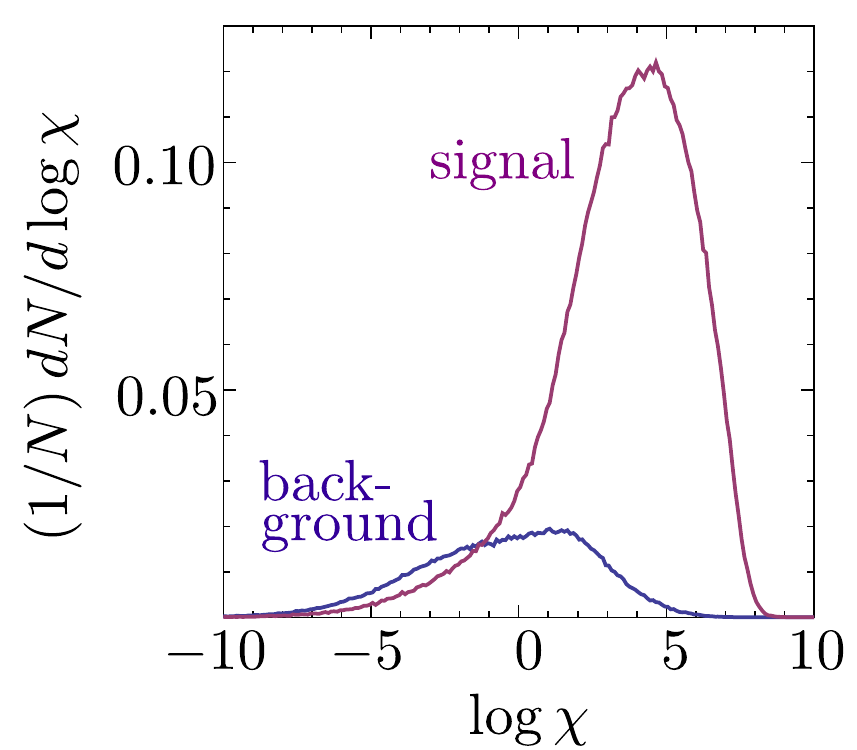}}
\caption{
$(1/N)\, dN/ d\log\chi$ for signal events (upper curve) and $(1/N)\, dN/ d\log\chi$ for background events (lower curve) for samples of signal and background events generated by \textsc{Pythia}. We use the cuts described in Sec.~\ref{sec:EventSelection}.
}
\label{fig:SandBvschi}
\end{figure}
%-------------END FIGURE----------------

\subsection{Distributions versus $\chi$}
\label{sec:chidistribution}

Using shower deconstruction, for each fat jet in the event, we calculate $\chi$. About 32\% of signal jets have $\chi = 0$ because the shower deconstruction algorithm cannot find a shower history that matches the signal hypothesis within the cuts that are built into the algorithm. This represents a failing suggesting that the algorithm is overly strict. However, 68\% of the signal jets remain. This number can be increased by increasing the top and $W$ mass windows. The distribution of the $\log\chi$ values for $\log\chi > -10$ is shown in the upper curve in figure \ref{fig:SandBvschi}. The bin with with $\chi = 0$ is not displayed. The distribution is normalized to the total number of generated signal jets, so that the integral under the curve including the $\chi = 0$ bin is 1 and the integral above $\log\chi = -10$ is about 0.68.

Similarly, for each generated background jet, we calculate $\chi$. About 86\% of these jets have $\chi = 0$ because the shower deconstruction algorithm cannot find a shower history that matches the signal hypothesis within the cuts that are built into the algorithm. That is, most of the background jets do not look at all like signal jets. About 14\% of the background jets remain. The distribution of the $\log\chi$ values for $\log\chi > -10$ is shown in the lower curve in Fig.~\ref{fig:SandBvschi}. The distribution is normalized to the total number of generated background jets, so that the integral under the curve including the $\chi = 0$ bin is 1 and the integral above $\log\chi = -10$ is about 0.12.

The idea of shower deconstruction is that the distribution in $\log\chi$ for signal jets should be very different from the distribution for background jets. We see in Fig.~\ref{fig:SandBvschi} that this is the case. First of all, most of the background jets have $\log\chi = - \infty$ and are not visible in the graph. Second, few background jets have $\chi > e^2$. On the other hand,  signal jets frequently have $\chi \approx e^4$.

\subsection{Discriminating signal from background}
\label{sec:Brejection}

\begin{figure}
\includegraphics[width=8.0cm]{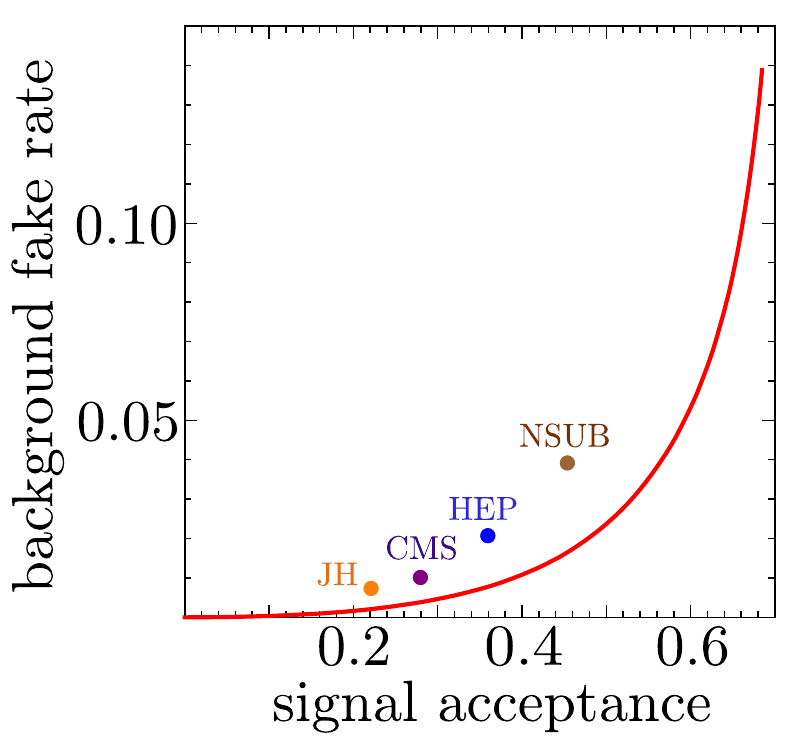}\ \
\includegraphics[width=8.0cm]{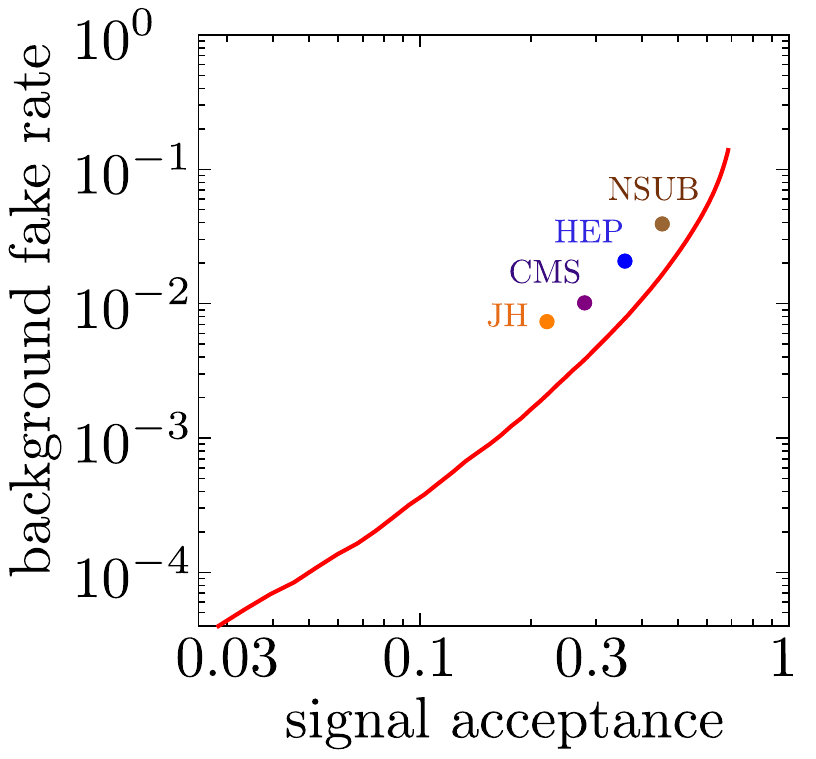}
\caption{
Background fake rate $F$ as a function of signal acceptance $A$ for shower deconstruction with the signal and background event samples described in Sec.~\ref{sec:EventSelection}. The curve for shower deconstruction is compared to $F$ vs $A$ points for the Johns Hopkins top tagger (JH), the top tagger of the CMS group (CMS), the Heidelberg-Eugene-Paris top tagger (HEP), and the use of N-subjettiness as a top tagger (NSUB). We show the results on a linear scale (left) and on a logarithmic scale (right).
}
\label{fig:acceptanceL}
\end{figure}

The simplest way to make use of the differing $\chi$ distributions between signal jets and background jets is to tag the fat jet as {\tt top} or {\tt other} according to whether $\chi$ is greater than or less than some fixed value $\chi_{\rm cut}$. With such a cut, some fraction $A$ of the signal jets will be correctly labeled as top jets. One calls $A$ the signal acceptance (or the tagging efficiency). Correspondingly, some fraction $F$ of the background jets will be incorrectly labeled as top jets. One calls $F$ the background fake rate (or the misstag rate). We want $A$ to be big and $F$ to be small. If we lower $\chi_{\rm cut}$, we make $A$ bigger, but unfortunately $F$ gets bigger at the same time. We can make $F$ smaller by raising $\chi_{\rm cut}$, but then $A$ gets smaller. This is illustrated in Fig.~\ref{fig:acceptanceL}.

%---------------FIGURE------------------
%-------------END FIGURE----------------

There are a number of available algorithms for tagging top jets. We compare shower deconstruction with the publicly available taggers mentioned in Sec.~\ref{sec:EventSelection}. We have used each of these in turn to tag the jets in our signal and background event samples, using the default parameters of the algorithm. For a specific choice of parameters the tagging performance can be characterized by one point on the signal acceptance {\it vs.}\ background fake rate plane. We have plotted the corresponding points in Fig.~\ref{fig:acceptanceL}. Notice that we use fixed windows in top mass and W mass and use the default parameters for each tagger. Then each tagger appears as a point in Fig.~\ref{fig:acceptanceL}. See Ref.~\cite{Altheimer:2012mn} for graphs in which the mass windows and input parameters are varied.

Using only fixed $m_\Lt$ and $m_\LW$ windows and the default input parameters, there is no definite answer to the question of which top tagger does the best job because each has a different signal acceptance. One might favor the HEP tagger over the JH tagger because the HEP tagger has a higher signal acceptance or might favor the JH tagger over the HEP tagger because the JH tagger has a lower background fake rate. Nevertheless, for any given signal acceptance $A$, a lower background fake rate $F$ is best. The ratio of the background fake rate $F_{\rm JH}$ for the JH tagger to the background fake rate $F_{\rm sd}(A_{JH})$ from shower deconstruction at the same signal acceptance $A_{\rm JH}$ as given by the JH tagger is about 3.6. Similarly, $F_{\rm CMS}/F_{\rm sd}(A_{\rm CMS}) \approx 2.7$, $F_{\rm HEP}/F_{\rm sd}(A_{\rm HEP}) \approx 2.6$, and $F_{\rm NSUB}/F_{\rm sd}(A_{\rm NSUB}) \approx 2.4$. For this reason, one may regard shower deconstruction as doing better than any of the previously available top taggers. The right plot of Fig.~\ref{fig:acceptanceL} shows the results on a logarithmic scale. With this plot, it is easier to see that one can gain a lot in making the background fake rate smaller if one is willing to sacrifice signal acceptance. For instance, with a signal acceptance of 0.1 one can reduce the background fake rate to about $5\times 10^{-4}$.

\subsection{Results with Herwig++}
\label{sec:herwig}

\begin{figure}
\centerline{\includegraphics[width=8.0cm]{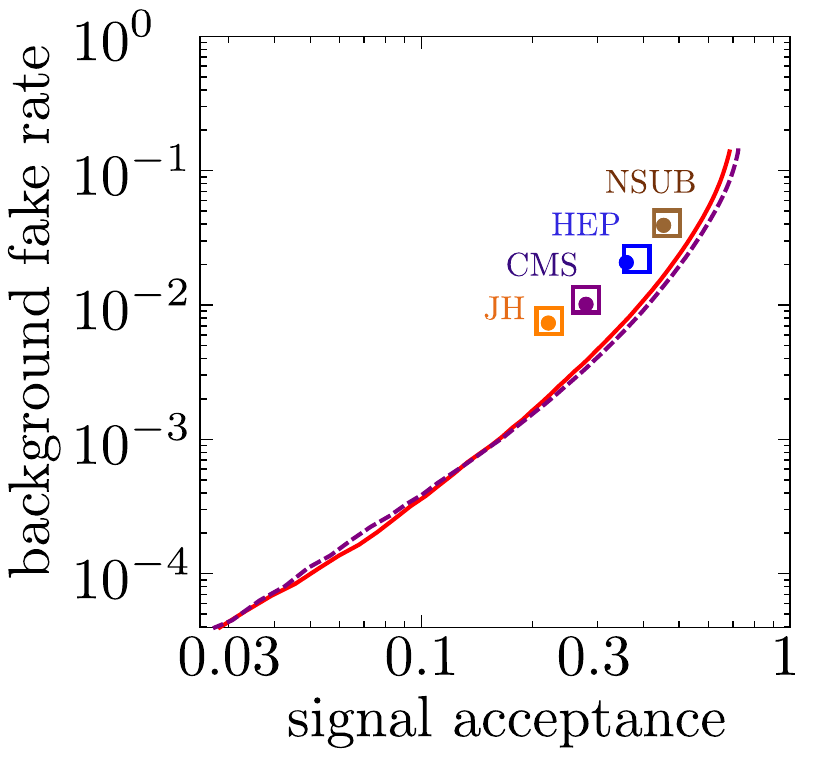}}
\caption{
Results using \textsc{Herwig++} compared to those using \textsc{Pythia} from Fig.~\ref{fig:acceptanceL}. The solid curve is the $F$ versus $A$ curve from shower deconstruction using events generated with \textsc{Pythia}; the dashed curve uses events generated with \textsc{Herwig++}. The solid circles show $F$ versus $A$ results for the top taggers using events generated with \textsc{Pythia}; the open squares use events generated with \textsc{Herwig++}.
}
\label{fig:acceptanceloglogHerwig}
\end{figure}

The results  presented above were based on signal and background events generated with \textsc{Pythia}. One may wonder whether these results are sensitive to which Monte Carlo event generator is used to generate events. To answer this question, we repeated the analysis using events generated with \textsc{Herwig++}. We find that with \textsc{Herwig++} signal events, $(1/N)\,dN/d\log\chi$ in the region $\chi > 0$ is about 8\% larger than with \textsc{Pythia} events, while $(1/N)\,dN/d\log\chi$ for background events generated with \textsc{Herwig++} is close to $(1/N)\,dN/d\log\chi$ for background events generated with \textsc{Pythia}. This leads to very similar results for the background fake rate as a function of signal acceptance. We display this comparison in Fig.~\ref{fig:acceptanceloglogHerwig}.

%---------------FIGURE------------------
%-------------END FIGURE----------------

\section{Results for low-$p_T$ top jet}
\label{sec:resultslowPT}

While the medium $p_T$ region of boosted top quarks, $\mathcal{O}(500)$ GeV, is a scenario most of the taggers we compare to are designed for, reconstructing top quarks with only a small boost, $\mathcal{O}(200)$ GeV, is more challenging. However, reconstructing top quarks in this low-$p_T$ region is phenomenologically highly relevant for a large variety of standard model \cite{HEPtagger,Kling:2012up} and beyond the standard model \cite{Kribs:2010ii,Hewett:2011wz,Hook:2012fd,Plehn:2012pr,Kaplan:2012gd} searches. 

Due to the smaller boost of the top quark, the decay products are widely separated. If the fat jet radius is not large enough to capture most of the decay products of the top quark, the taggers will not be able to positively identify a top jet. Therefore, a large cone size is necessary to reconstruct top quarks with small boost. However, this will allow a lot of top-uncorrelated radiation to enter the fat jet, {\it i.e.} initial state radiation and contributions from the underlying event.

Compared to the scenario studied in Sec.~\ref{sec:results}, we only change the fat jet algorithm and the related event selection cuts. We reconstruct the fat jets using again the Cambridge/Aachen algorithm but now with R=1.5. Events are accepted for further analysis if they provide at least two jets with $p_{T,j} \geq 200$ GeV each.

\begin{figure}
\centerline{\includegraphics[width=8.0cm]{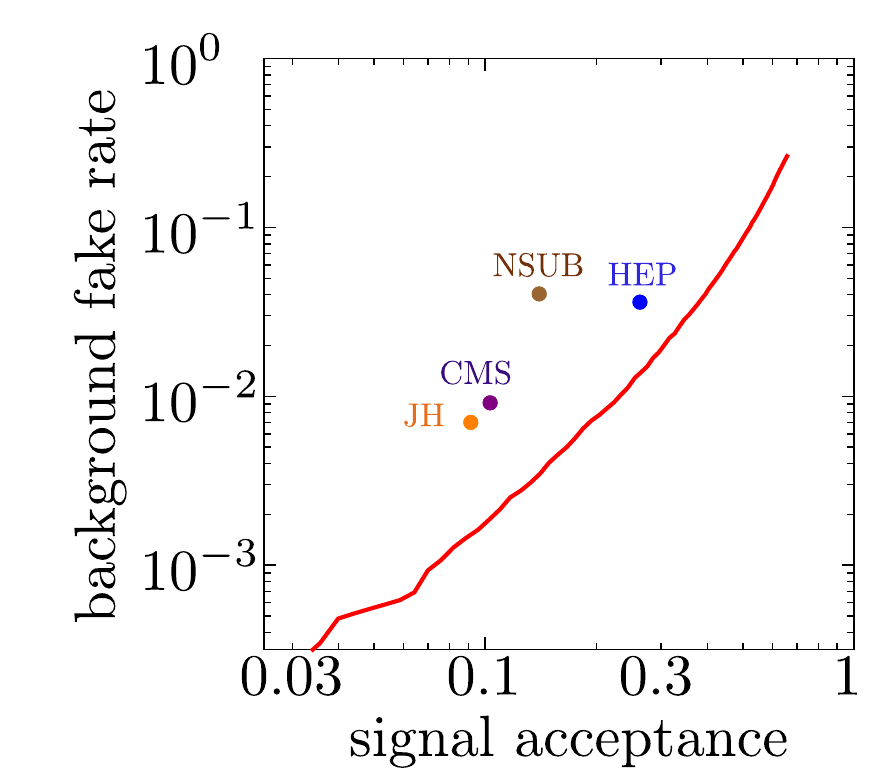}}
\caption{
Background fake rate $F$ as a function of signal acceptance $A$ for shower deconstruction with the signal and background event samples using a $200\ \text{GeV}$ cut on jet $P_T$ and a fat jet cone size of R=1.5, as described in Sec.~\ref{sec:resultslowPT}. The curve for shower deconstruction is compared to $F$ vs $A$ points for the Johns Hopkins top tagger (JH), the top tagger of the CMS group (CMS), the Heidelberg-Eugene-Paris top tagger (HEP), and the use of N-subjettiness as a top tagger (NSUB).
}
\label{fig:lowptresult}
\end{figure}

We find that all taggers perform worse in this scenario compared to Sec.~\ref{sec:results}; see Fig. \ref{fig:lowptresult}. However, even in this challenging scenario shower deconstruction performs better than the other taggers. Here the relative improvements are $F_{\rm{JH}}/F_{\rm{sd}}(A_{\rm{JH}}) \sim 4.2$, $F_{\rm{CMS}}/F_{\rm{sd}}(A_{\rm{CMS}}) \sim 4.6$, $F_{\rm{HEP}}/F_{\rm{sd}}(A_{\rm{HEP}}) \sim 2.6$, $F_{\rm{NSUB}}/F_{\rm{sd}}(A_{\rm{NSUB}}) \sim 11.9$. The HEPTopTagger is the only tagger explicitly designed to work for low $p_T$ top quarks. Consequently it shows the smallest change in the performance ratio compared to the scenario with medium boosted top quarks. Crucial for a good performance in this scenario is a built-in grooming procedure which the CMS and JH tagger largely and the NSubjettiness tagger completely lack. Thus, particularly for the NSubjettiness tagger, one can expect a performance improvement by changing the top mass window.

%---------------FIGURE------------------
%-------------END FIGURE----------------

\section{Cone size dependence}
\label{sec:resultsdiffR}

\begin{figure}
\centerline{\includegraphics[width=8.0cm]{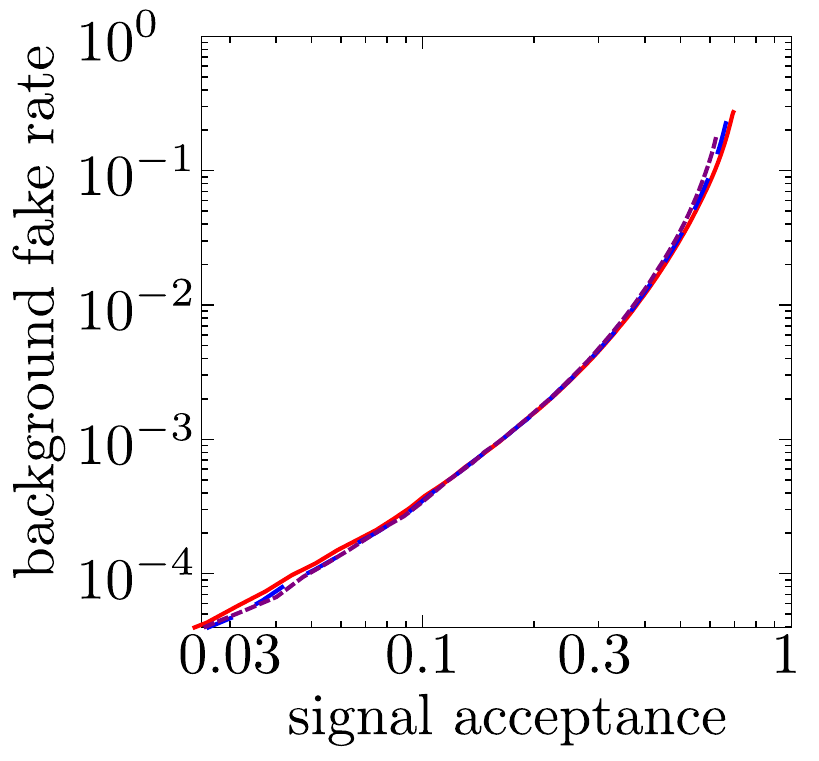}}
\caption{
Shower deconstruction results for highly boosted jets ($P_T > 800\ \text{GeV}$) showing the dependence on the cone size of the fat jet. The solid curve is the $F$ versus $A$ curve from shower deconstruction for fat jets defined with $R = 1.50$; the long dashed curve uses jets with $R = 1.25$;  the short dashed curve uses jets with $R = 1.00$.
}
\label{fig:acceptancelogloghighT}
\end{figure}

In this section, we study how sensitive shower deconstruction is with respect to the cone size and the overall amount of uncorrelated soft radiation in the fat jet. We use  an event sample in which the fat jet is highly boosted: we require $P_T > 800\ \text{GeV}$. We then plot in Fig.~\ref{fig:acceptancelogloghighT} the background fake rate versus signal acceptance for cone sizes $R = 1.5$, 1.25, and 1.00. We see that the cone size makes very little difference. The larger cone sizes include more debris from initial state radiation, but the shower deconstruction algorithm seems not to be confused by this debris.

%---------------FIGURE------------------
%-------------END FIGURE----------------

\section{Measuring parameters of the theory}
\label{sec:Wmasshypothesis}

In many applications of shower deconstruction, some parameters of the theory for the sought signal events may not be known. In that case, one would like not only to show from the data that the sought signal is present in nature but one would also like to measure the unknown parameters. In the example used in this paper, suppose that we did not know the mass $M_\LW$ of the W boson. Then we could find $M_\LW$ from the data. There is one true $M_\LW$ in nature (80.4 GeV in our Monte Carlo event sample). However $M_\LW$ is also a parameter in the model used in the shower deconstruction algorithm. If the model $M_\LW$ is not right, then the shower deconstruction results should tell us.

In a complete analysis, one would construct from the data the ratio of the likelihood that the observed data is generated by the signal plus the QCD background to the likelihood that the data is generated by background only. Then this likelihood ratio should be small if the model $M_\LW$ is far from the true $M_\LW$ and should peak at $M_\LW^{\rm model} = M_\LW$.

\begin{figure}
\centerline{\includegraphics[width=8.0cm]{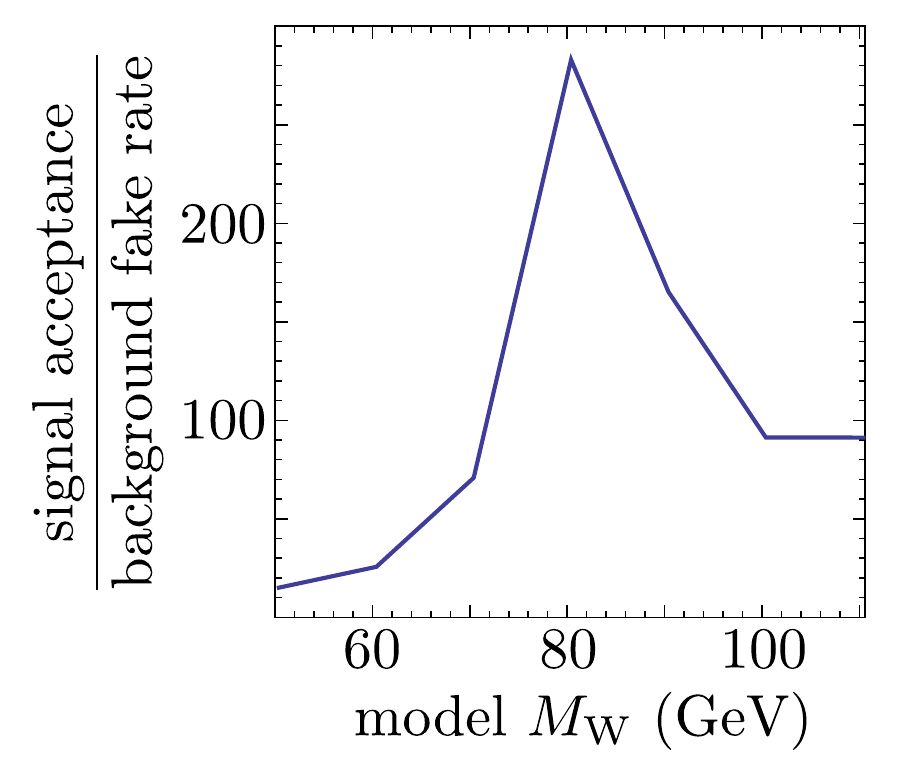}}
\caption{
Signal acceptance divided by background fake rate for a cut $\chi > 384$ as a function of the W mass, $M_\LW^{\rm model}$, used in the shower deconstruction algorithm. 
There is a peak at the true W mass.}
\label{fig:wmass}
\end{figure}

To explore this with a simpler calculation, we have, for the event sample described in Sec.~\ref{sec:EventSelection}, applied shower deconstruction for a range of model $M_\LW$ choices. Then we have calculated the background fake rate and the signal acceptance with the cut $\chi > 384$, which corresponds to approximately a 20\% signal acceptance when $M_\LW^{\rm model} = M_\LW$. The background fake rate rises slowly as $M_\LW^{\rm model}$ increases. The signal acceptance has a peak at $M_\LW^{\rm model} = M_\LW$. We calculated the ratio of signal acceptance to background fake rate as a function of $M_\LW^{\rm model}$. The results are shown in Fig.~\ref{fig:wmass}. We see that this ratio, as expected, exhibits a peak at $M_\LW^{\rm model} = M_\LW$. We notice that the shape of the curve is not symmetric: a real signal event can look like a $M_\LW^{\rm model} > M_\LW$ signal event when extraneous gluons from initial state radiation get counted as part of the W decay products.

%---------------FIGURE------------------
%-------------END FIGURE----------------

\section{Conclusion and prospects}
\label{sec:conclusion}

In this paper, we have developed an algorithm for tagging top jets based on the method of shower deconstruction. For this purpose we had to considerably extend the shower deconstruction approach designed to reconstruct a Higgs boson as outlined in Ref.~\cite{SSI}. The approach models parton evolution from the hard interaction scale at which a boosted top quark is created down to the virtuality scale of the microjets that serve as the input to the calculation. For this, one needs the decay matrix elements for $\Lt \to \LW + \Lb$ and then $\LW \to q + \bar q$. Then one needs the splitting probabilities and Sudakov factors for QCD showering for the massive top quark, a massless bottom quark, and light partons created in the W boson decay. The splitting probabilities include appropriate factors for quantum interference for radiation of soft gluons from color dipoles.

We find that shower deconstruction performs significantly better than any of the publicly available taggers that we compared with for either a moderately boosted top quark with $P^\text{jet}_T > 500\ \text{GeV}$ or one that is only boosted to $P^\text{jet}_T > 200\ \text{GeV}$. Also, we found that the performance of shower deconstruction is not very sensitive to the cone size used to define the fat jet as long as the cone size is large enough to contain the top quark decay products. 

Because shower deconstruction performs a hypothesis test for competing theories or processes it can be used to measure their parameters. As an example we varied the W boson mass in the reconstruction algorithm of the top. When the hypothesis matched nature, as simulated by a full event generator, the reconstruction significance was maximized, thereby allowing to measure the W boson's mass.

Our subject in this paper has been limited to distinguishing top quark jets from background jets. One can also imagine assigning a variable $\chi$ to events containing multiple jets according to the ratio of the likelihoods that the event was produced by a signal process of interest or was  produced by an ordinary background process. For instance, one could look for events produced by the decay of a new, heavy, vector boson $\LZ'$ that decays to ${\rm t} + \bar{\rm t}$. Then we need to distinguish such signal events from Standard Model events with two jets that may, or may not, be top jets. Shower deconstruction of individual jets can, we believe, be extended to cover event deconstruction of whole events. We leave this extension to future work.

\acknowledgments{ 
This work was supported by the United States Department of Energy. We thank Daniel Steck and Jeremy Thorn for the use of the Quantum Control computer cluster at the University of Oregon. We are grateful to Z.~Nagy for advice and for his consent to incorporate some as yet unpublished parts of his work on the construction of parton showers into this work. We thank B.~Webber for helpful conversations about how parton showers and decaying particles fit together.}

%==========================================================================

\appendix
\section{Appendix}
\label{sec:appendix}

In this appendix, we fill in some of the details about the factors that go into shower deconstruction for this analysis. Most of the ingredients are the same as in Ref.~\cite{SSI}. Thus we present only new features that are needed to include the decays of top quarks and W bosons and to include gluon radiation from the top quark.

\subsection{Top quark decay probability}

In a parton shower, the total probability for a splitting has the form $H e^{-S}$, where $H$ is the probability that the parton splits at shower time $t$ and $e^{-S}$ is the probability that it has not split at an earlier shower time. We can formulate parton decay in the same way. For the decay, let us denote $H e^{-S} = \widetilde H$. Then for a top quark decay, we take
\begin{equation}
\label{eq:tildeHform}
\widetilde H_\Lt = C_\Lt
\frac{2\pi m_\Lt\Gamma_\Lt}{2\arctan(\Delta_\Lt/\Gamma_\Lt)}
\frac{\Theta(|p_t^2 - m_\Lt^2| < m_\Lt\Delta_\Lt)}
{(p_t^2 - m_\Lt^2)^2 + m_\Lt^2 \Gamma_\Lt^2}
\;,
\end{equation}
where
\begin{equation}
C_\Lt = \frac{8\pi m_\Lt^2}{m_\Lt^2 - m_\LW^2}
\;.
\end{equation}
The main feature of this is the standard Breit-Wigner denominator, $(p_t^2 - m_\Lt^2)^2 + m_\Lt^2 \Gamma_\Lt^2$, where $\Gamma_\Lt$ is the decay width.\footnote{We choose the simulated width $\Gamma_\Lt$ larger than the physical top quark width in order to approximately simulate an imperfect resolution in measuring jet momenta. See Sec.~\ref{sec:taggerparameters}.} For shower deconstruction, we supply an extra factor, $\Theta(|p_t^2 - m_\Lt^2| < M_\Lt\Delta_\Lt)$, where $\Delta_\Lt$ is greater than $\Gamma_\Lt$. We insert this factor as an approximation in order to eliminate entirely shower histories for which $\widetilde H$ would be small. There are two normalization factors. One is fixed by
\begin{equation}
\int\!\frac{dp_t^2}{2\pi}\
\frac{2\pi\, m_\Lt\Gamma_\Lt}{2\arctan(\Delta_\Lt/\Gamma_\Lt)}
\frac{\Theta(|p_t^2 - m_\Lt^2| < m_\Lt\Delta_\Lt)}
{(p_t^2 - m_\Lt^2)^2 + m_\Lt^2 \Gamma_\Lt^2}
=1
\;.
\end{equation}
The second, $C_\Lt$, is fixed by
\begin{equation}
C_\Lt\
(2\pi)^{-3}\int\! \frac{d\vec p_b}{2 \omega_b}\
(2\pi)^{-3}\int\! \frac{d\vec p_W}{2 \omega_W}\
(2\pi)^4\delta(p_b + p_W - p_t)
= 1
\end{equation}
as long as $p^2_t = m_\Lt^2$ and $p^2_\LW = m_\LW^2$ are good approximations. Together, these normalization factors insure that the top quark decays with probability 1.

Now we need the Sudakov exponent $S_\Lt(t,t_0)$, which is the integral of $\widetilde H/C_\Lt$ from a starting shower time $t_0$ defined by the previous splitting to a shower time $t$ related to $|p_t^2 - m_\Lt^2|$ according to Eq.~(\ref{eq:showertimedecaydef}). If we define
\begin{equation}
\label{eq:tmindef}
t_{\rm min} = \log\!\left(\frac{|Q_0| k_t}{m_\Lt\Delta_\Lt}\right)
\;,
\end{equation}
then $S_\Lt(t,t_0) = S_\Lt(t,t_{\rm min}) - S_\Lt(t_0,t_{\rm min})$. Taking into account the jacobian to change integration variables from $p_t$ to $t$, we get
\begin{equation}
\begin{split}
\label{eq:tdecaySudakovexonent}
S_\Lt(t,t_{\rm min})
={}& \log\left[\arctan(\Delta_\Lt/\Gamma_\Lt)\right]
- \log\left[
\arctan\!\left(\frac{|Q_0| k_t}{m_\Lt \Gamma_\Lt}\,
e^{-\max(t,t_{\rm min})}\right)\right]
\;.
\end{split}
\end{equation}
Having found $S_\Lt$, we immediately obtain the decay function without the Sudakov factor,
\begin{equation}
\label{eq:Hform}
H_\Lt = C_\Lt\,
\frac{2\pi m_\Lt\Gamma_\Lt}{2\arctan(|p_t^2 - m_\Lt^2|/[m_\Lt\Gamma_\Lt])}\,
\frac{\Theta(|p_t^2 - m_\Lt^2| < m_\Lt\Delta_\Lt)}
{(p_t^2 - m_\Lt^2)^2 + m_\Lt^2 \Gamma_\Lt^2}
\;.
\end{equation}

We have so far considered top quark decay in isolation. However, a top quark carries color and therefore can emit a gluon. In any interval $dt$ of shower time, the top quark can either emit a gluon or decay. The top quark emits a gluon with a probability determined by a splitting function $H_{\Lt \Lt \Lg}$ that we will discuss in section~\ref{sec:topsplitting}. The gluon emission process has its own Sudakov exponent, $S_{\Lt \Lt \Lg}$. The probability that the top quark has neither emitted a gluon nor decayed by shower time $t$ is given by the sum of the Sudakov exponents, $S_\Lt + S_{\Lt \Lt \Lg}$.

\subsection{W boson decay probability}

The W boson created in the top quark decay will itself decay to a quark $q$ and an antiquark $\bar q$. For the total splitting probability $\widetilde H = H e^{-S}$, we take
\begin{equation}
\label{eq:tildeHformW}
\widetilde H_\LW = 8\pi\
\frac{2\pi\, m_\LW\Gamma_\LW}{2\arctan(\Delta_\LW/\Gamma_\LW)}
\frac{\Theta(|p_\LW^2 - m_\LW^2| < m_\LW\Delta_\LW)}
{(p_\LW^2 - m_\LW^2)^2 + m_\LW^2 \Gamma_\LW^2}\
g_\LW(p_q,p_{\bar q},p_t)
\;.
\end{equation}
This is like the decay probability for the top quark except that now we have an extra function $g$. The W has spin 1 and it is polarized. That is, it has a non-trivial spin density matrix. That happens because of the decay process that created the W. The W-polarization leads to an angular dependence of the decay products' momenta as seen in the W rest frame. This angular dependence is represented by the function $g_\LW$. Since the polarization arises from the top decay, $g_\LW$ depends on $p_t$. Specifically,
\begin{equation}
g_\LW = \frac{12\, p_{\bar q}\cdot p_t (p_t - p_{\bar q})^2}
{(m_\Lt^2 - m_\LW^2)(m_\Lt^2 + 2 m_\LW^2)}
\;.
\end{equation}

Since the W boson is colorless, there is not a competition between W decay and gluon emission. For this reason, it is enough to represent the total probability for the W to decay by $\widetilde H_\LW$ without separately using a Sudakov factor.

\subsection{Top quark splitting function}
\label{sec:topsplitting}
The top quark can emit a gluon. The splitting function for this, $H_{\Lt \Lt \Lg}$, differs from a light quark splitting function because of the top quark mass. Following closely the reasoning in Ref.~\cite{SSI}, we take
\begin{equation}
\begin{split}
\label{eq:splittingttg}
H_{\Lt \Lt \Lg} ={}&  
\frac{8\pi C_\LF\as}{\mu_J^2} \,
\frac{k_J}{k_g}
\left[ 1 + \left(\frac{k_t}{k_J}\right)^{\!2}\right]
g(p_g,p_t,p_k)\,
\Theta\!\left(
2\,\frac{\mu_J^2}{k_J}
< \frac{\mu_K^2}{k_K}
\right)
\;.
\end{split}
\end{equation}
Here $J$ refers to the mother top quark and $\mu_J^2 = p_J^2 - m_\Lt^2$. Then $t$ and $g$ refer to the daughter top quark and the daughter gluon, respectively, and $k_t$ and $k_g$ are their transverse momenta. If we denote $k_g \approx (1-z) k_J$ and $k_t \approx z k_J$, we recognize the familiar collinear splitting function $(1 + z^2)/(1-z)$ in $H_{\Lt \Lt \Lg}$. There is a theta function that enforces ordering of the shower emissions in shower time. In this theta function, $\mu_K^2$ denotes the virtuality in the previous emission from the top quark and $k_K$ denotes the transverse momentum of the top quark just before this emission.  In a strongly ordered shower, we would have ${\mu_J^2}/{k_J} \ll  {\mu_K^2}/{k_K}$. In our shower, we settle for a factor of 2 between these scales. In the case that there was no previous splitting, the theta function in $\widetilde H$ is not needed and we ignore it, while in the corresponding calculation of the Sudakov exponent we replace ${\mu_K^2}/{k_K} \to  2 (k_J^2 + m_\Lt^2)/k_J$ in the theta function.

When the emitted gluon is soft, there is quantum interference between emission of the gluon from the top quark and emission from some other (massless) parton $k$ that is color connected to the top quark. We partition the emission probability from the whole dipole into two terms, one of which looks mostly like emission from the top quark and the other of which looks mostly like emission from parton $k$. The term that looks mostly like emission from the top quark is $H_{\Lt \Lt \Lg}$. The influence of the color connected partner is seen in the function $g(p_g,p_t,p_k)$, which is
\begin{equation}
\label{eq:softsplittingTop}
g(p_g,p_t,p_k) =   \frac{k_g\, p_g \cdot p_t}{2 k_t}\
\frac{- (p_g\cdot p_t\ p_k - p_g\cdot p_k\ p_t)^2}
{(p_g \cdot p_t\  p_g \cdot p_k)^2}\
A_{tk}'
\;.
\end{equation}
The first factor here is simply the inverse of the soft gluon limit of the factors that we have included in the collinear part of $H_{\Lt \Lt \Lg}$. The second factor is the squared matrix element for emission of a soft gluon with momentum $p_g$ from a dipole consisting of partons with momenta $p_t$ and $p_k$. The third function is a function $A_{tk}'$ that serves to partition the dipole squared matrix element into the two terms mentioned above. There is some arbitrariness in choosing this function. As in Ref.~\cite{SSI}, we take the choice given in Eq.~(7.12) of Ref.~\cite{NSIII},
\begin{equation}
A_{tk}' = \frac{p_g\cdot p_k\ k_t}
{p_g\cdot p_k\ k_t + p_g\cdot p_t\ k_k}
\;.
\end{equation}
After expanding the factors here, we have
\begin{equation}
g(p_g,p_t,p_k)
=
\frac{k_g}
{2\,p_g \cdot p_t}
\frac{ 2\, p_g\cdot p_t\  p_t\cdot p_k
- m_\Lt^2 \, p_g\cdot p_k}
{p_g\cdot p_k\ k_t + p_g\cdot p_t\ k_k}
\;.
\end{equation}
It is convenient to write this in terms of the angles between the partons, using the approximation that these angles are small. Using rapidities $y$ and azimuthal angles $\phi$ of the partons, we define
\begin{equation}
\begin{split}
\theta^2_{gt} ={}& (y_g - y_t)^2 + (\phi_g - \phi_t)^2
\;,
\\
\theta^2_{gk} ={}& (y_g - y_k)^2 + (\phi_g - \phi_k)^2
\;,
\\
\theta^2_{tk} ={}& (y_t - y_k)^2 + (\phi_t - \phi_k)^2
\;.
\end{split}
\end{equation}
Then for small angles and $m_\Lt^2/k_t^2 \ll 1$ we have the function $g$ in the form in which we use it to compute $H_{\Lt \Lt \Lg}$:
\begin{equation}
\label{eq:gsmallfull}
g(p_g,p_t,p_k)
=
\frac{ (\theta^2_{gt} + m_\Lt^2/k_t^2)  (\theta^2_{tk} + m_\Lt^2/k_t^2)
- (m_\Lt^2/k_t^2) \, \theta^2_{gk}}
{(\theta^2_{gt} + m_\Lt^2/k_t^2)(\theta^2_{gk}  + \theta^2_{gt} + m_\Lt^2/k_t^2)}
\;.
\end{equation}
Notice that, by construction, $g$ is {\em not} singular when $\theta^2_{gk} \to 0$.

\subsection{Massless parton splitting functions}

We treat all quarks except for the top quark as being massless. When a massless quark splits by emitting a gluon, the splitting function is
\begin{equation}
\begin{split}
\label{eq:splittingHqqg}
H_{qqg} ={}&  
\frac{8\pi C_\LF\as}{\mu_J^2} \,
\frac{k_J}{k_g}
\left[ 1 + \left(\frac{k_q}{k_J}\right)^2\right]
g(p_g,p_q,p_k)\,
\Theta\!\left(
2\,\frac{\mu_J^2}{k_J}
< \frac{\mu_K^2}{k_K}
\right)
\;.
\end{split}
\end{equation}
Here $J$ refers to the mother quark and $\mu_J^2 = p_J^2$. Then $q$ and $g$ refer to the daughter quark and the daughter gluon, respectively, and $k_q$ and $k_g$ are their transverse momenta. When a gluon splits by emitting a gluon, the splitting function is
\begin{equation}
\begin{split}
\label{eq:splitingHggg}
H_{ggg} ={}&  
\frac{8\pi C_\LA\as}{\mu_J^2} \,
\frac{k_J^2}{k_s k_h}
\left[ 1 - \frac{k_s k_h}{k_J^2}\right]^2
g(p_s,p_h,p_k)\,
\Theta\!\left(
2\,\frac{\mu_J^2}{k_J}
< \frac{\mu_K^2}{k_K}
\right)
\;.
\end{split}
\end{equation}
Now $h$ and $s$ refer to the daughter gluon with the greater transverse momentum $k_h$ and the daughter gluon with the smaller transverse momentum $k_s$, respectively. For the gluon splitting, if we approximate $k_s/k_J = 1-z$ and $k_h/k_J = z$, we see that $H_{ggg}$ contains a collinear splitting factor $[1-z(1-z)]^2/[z(1-z)]$, in contrast to the quark splitting factor $[1+z^2]/(1-z)$. In both $H_{qqg}$ and $H_{ggg}$, there is a theta function that enforces ordering of the shower splittings in shower time, as in the previous subsection. Except for the function $g$, these are the same functions $H$ that we used in Ref.~\cite{SSI}.

There is also a function $g$. When the emitted gluon is soft, there is quantum interference between emission of the gluon from parton $J$ and emission from some other parton $k$ that is color connected to the splitting parton. We partition the emission probability from the whole dipole into two terms, as in the previous subsection. The influence of the color connected partner is seen in the function $g$. This is the same function for emission from a quark and emission from a gluon, but with different variable names. With the same logic as in the previous section, we have
\begin{equation}
\begin{split}
\label{eq:gdefquark2}
g(p_g,p_q,p_k) ={}& 
\frac{(\theta^2_{gk} + m_k^2/k_k^2) (\theta^2_{qk} + m_k^2/k_k^2)
- (m_k^2/k_k^2) \,\theta^2_{gq}}
{(\theta^2_{gk} + m_k^2/k_k^2)(\theta^2_{gq} + \theta^2_{gk} + m_k^2/k_k^2)}
\;.
\end{split}
\end{equation}
This is the same function that we used in Ref.~\cite{SSI} except that here the color connected partner $k$ could be massive because it could be the top quark.

\subsection{Dipole antenna splitting}

In shower II, a massless parton can emit a gluon with the participation of a color connected parton that is the top quark just before its decay. In this case, a color dipole emits the gluon and we do not partition the emission into two pieces. Rather, we consider the dipole to be a unit, sometimes called a dipole antenna. The splitting function is then given by Eq.~(\ref{eq:splittingHqqg}) or Eq.~(\ref{eq:splitingHggg}), depending on whether the emitting parton is a quark or a gluon. The only difference with the preceding section is that now we omit the partitioning function $A'_{qk}$ or $A'_{hk}$. With this choice, the angular function is
\begin{equation}
\begin{split}
\label{eq:gprimedefquark2}
g(p_g,p_q,p_k) ={}& 
\frac{(\theta^2_{gk} + m_k^2/k_k^2) (\theta^2_{qk} + m_k^2/k_k^2)
- (m_k^2/k_k^2) \,\theta^2_{gq}}
{(\theta^2_{gk} + m_k^2/k_k^2)^2}
\;.
\end{split}
\end{equation}
Here parton $k$ is the top quark, so $m_k = m_\Lt$. Notice that if we were to set $m_k$ to zero, this function would be singular when $\theta^2_{gk} \to 0$. That is the consequence of omitting $A'_{qk}$. Because $m_\Lt > 0$, there is no singularity.
 
\subsection{Sudakov exponents}

For each propagator in a shower history diagram, there is a Sudakov factor $e^{-S}$. This factor gives the probability for the parton not to have split between the shower time of its previous splitting and the shower time of the next splitting. If there is no next splitting, then $e^{-S}$ represents the probability not to have split between the previous splitting and the shower time that corresponds to the microjet virtuality. The top quark can either split or decay, so there are two contributions to $S$. The W boson can only decay, so we simply include $e^{-S}$ in the function $\widetilde H$ that gives the differential decay probability.

We calculate Sudakov exponents for QCD splittings using
\begin{equation}
\label{eq:SudakovExponentdef}
S = \frac{1}{4(2\pi)^3}
\int\!d\mu_J^2 \,\Theta(\mu_{\rm min}^2 < \mu_J^2)
\int\!dz \int\!d\varphi\
H(\bar p_a,\bar p_g)\,
\;.
\end{equation}
Here $\mu_{\rm min}^2$ is the virtuality of the parton splitting. There is a $\mu_{\rm max}^2$ corresponding to the shower time of the previous splitting. The constraint $\mu_J^2 < \mu_{\rm max}^2$ is included in the splitting function $H$. The splitting functions $H$ are given in Ref.~\cite{SSI} and in the preceding subsections. The variable $z$ is the momentum fraction of the splitting and $\varphi$ is the azimuthal angle of the plane of the splitting about the direction of the mother parton.

We need to express $S$ as a quickly computable function of the variables in the shower history. Thus we cannot use numerical integration to evaluate the integrals in the definition (\ref{eq:SudakovExponentdef}). On the other hand, the integrals are too complicated to evaluate analytically. For that reason, we have developed simple numerical approximations to the integrals and we use these approximate functions. The approximations used are not really an essential part of the physics: the ones that we use currently are different from those used in Ref.~\cite{SSI} and if we found better approximations, we would use them. For that reason, it does not seem useful to list the approximate functions used to represent the functions $S$.

%-------------------------------------------------------------------
\end{document}